\documentclass[sigconf]{acmart}

\usepackage{bbm}

\AtBeginDocument{%
  }

\setcopyright{rightsretained}
\copyrightyear{2026}
\acmDOI{}
\acmConference[KDD '26]{5th Workshop on End-End Customer Journey Optimization at KDD 2026.}{August 09--13,2026}{Jeju Island, Republic of Korea}
\acmISBN{}

\begin{document}

\title{Session-Level Optimization for Large-Scale Retrieval using REINFORCE with Multi-Step Off-Policy Correction}

\author{Artem Matveev}
\email{matfu21@yandex.ru}
\orcid{0009-0004-0271-221X}
\affiliation{
  \institution{AI VK}
  \city{Moscow}
  \country{Russia}
}

\author{Sergei Makeev}
\email{neuralsrg@gmail.com}
\orcid{0009-0003-5451-6475}
\affiliation{
  \institution{AI VK}
  \city{Moscow}
  \country{Russia}
}

\author{Aleksei Krasilnikov}
\email{TheKabeton@yandex.ru}
\orcid{0009-0007-3484-7970}
\affiliation{
  \institution{AI VK}
  \city{Moscow}
  \country{Russia}
}

\author{Vladimir Baikalov}
\email{deadinside@itmo.ru}
\orcid{0009-0009-4864-2305}
\affiliation{
  \institution{AI VK}
  \city{Moscow}
  \country{Russia}
}

\author{Sergei Liamaev}
\email{liamaev.sergei@gmail.com}
\orcid{0009-0009-6316-1091}
\affiliation{
  \institution{AI VK}
  \city{Moscow}
  \country{Russia}
}

\author{Kirill Khrylchenko}
\email{elightelol@gmail.com}
\orcid{0009-0007-3640-8795}
\affiliation{%
  \institution{HSE University}
  \city{Moscow}
  \country{Russia}
}

\renewcommand{\shortauthors}{Matveev, Makeev et al.}

\begin{abstract}
Two-tower models are a widely used paradigm for large-scale retrieval in recommendation. However, they are typically trained with myopic supervised objectives, such as next-item prediction, that do not directly optimize long-term user satisfaction.

In this work, we formulate recommendation as a session-level sequential decision-making problem and train a two-tower retriever autoregressively with off-policy REINFORCE on pre-collected data. Unlike the one-step off-policy correction used in prior work, we propose a multi-step approximation of importance weights enabled by the autoregressive formulation.

To support offline evaluation, we train a user feedback model that simulates user responses to generated recommendations. This lets us adapt doubly robust off-policy evaluation for sequential decision-making to recommendation, a setting that has received limited attention. We further introduce a feedback-model-based test-time scaling procedure that simulates future responses and selects the recommendation with the highest predicted long-term return.

Experiments on the public large-scale Yambda-5B dataset show that our RL agent achieves higher off-policy estimates of cumulative session reward than next-item and next-positive prediction baselines, while remaining competitive on conventional retrieval metrics. Moreover, allocating more inference-time compute to simulating future responses yields higher model-based long-term returns without updating the policy.
\end{abstract}



\maketitle

\section{Introduction}
Recommender systems are widely deployed across industry to help users navigate vast collections of content. Their mission is to retain millions of users by presenting the most relevant items. To achieve this goal, recommender systems are typically optimized with respect to immediate feedback signals, such as clicks\,\cite{wang2020dcnv2} or repins\,\cite{xia2023transact}, which can be intuitively viewed as searching for a local optimum in the space of user interests.

\paragraph{Reinforcement learning}
Reinforcement learning (RL) offers a way to extend this optimization objective to account for long-term rewards. Intuitively, rather than recommending an item that appears locally optimal at a given instant, the recommender is tasked with selecting a sequence of actions that jointly maximize some notion of user satisfaction (reward). Despite the impressive advances in robotics\,\cite{kober2013reinforcement}, games\,\cite{silver2016mastering}, and autonomous driving\,\cite{kiran2021deep}, deployment of RL recommender system is challenging due to extremely sparse feedback\,\cite{ploshkin2025yambda}, uncertainty in reward design\,\cite{zhou2025onerecv2}, infrastructural requirements\,\cite{zhou2025onerec} and the limitations of offline evaluation\,\cite{chen2018topk}.

Depending on whether the agent is trained through interaction with the environment or using pre-collected data, RL can be divided into on-policy and off-policy methods, respectively. On-policy RL may be particularly challenging when low-latency infrastructure is required, which is often the case in online learning, where the deployed model is updated on-the-fly as new user feedback is received\,\cite{zhou2025onerec}. Moreover, existing on-policy deployments, such as OneRec\,\cite{zhou2025onerec} and OxygenREC\,\cite{hao2025oxygenrec}, in fact achieve consistency with the reward model (i.e., the ranker) rather than optimize long-term reward. Another on-policy RL application is the integration of personalized recommendation within language models, such as OneRec-Think\,\cite{liu2025onerecthink}. The RL stage there, however, is mostly designed to direct the language model's reasoning capabilities toward predicting the next interaction instead of optimizing long-term reward. Off-policy methods ease these requirements by utilizing pre-collected data (sampled from the behavior policy), thus enabling batch training.

Arguably, one of the most challenging parts of training an RL recommender is evaluation: how should one determine which of the learned policies is the best? While most works place this burden on A/B tests\,\cite{chen2018topk, chen2022offpolicy}, offline evaluation is undoubtedly crucial for rapid experimentation. Off-policy evaluation (OPE) can be thought of as a more sophisticated problem than off-policy learning itself, since the latter is expected to find a direction in which the policy improves, whereas the former must be able to evaluate a policy shifted in an arbitrary direction.

Reward models, which are often required by OPE methods, can also find use during inference. In large language models, test-time scaling is a process that iteratively improves generated outputs by scoring candidates with a reward model\,\cite{snell2024scaling, wang2022self}. Prior work has adapted test-time scaling to recommenders in the case of one-step trajectories for maximizing immediate reward\,\cite{guo2026promise}. In this work, we discuss the application of test-time scaling to the multi-step setting.

\paragraph{Retrieval}
Industrial recommenders often decouple the recommendation task into retrieval and ranking stages. Retrievers form a set of potentially relevant candidates, which are then sorted by a ranking model. There are two main reasons why a separate retrieval stage is needed. First, rankers are often slow and require significant compute, and in the case of pairwise ranking, where the ranker determines which of two candidates should be preferred, the search space becomes infeasibly large. Second, rankers suffer from the folding effect\,\cite{xin2017folding}, induced by the fact that they are trained on already relevant examples.

In recent years, two-tower retrievers with separate user and item encoders have become popular among industry practitioners\,\cite{yi2019sampling, yang2020mixed, pancha2022pinnerformer, khrylchenko2023personalized, khrylchenko2025argus}. This is largely because such models enable approximate nearest neighbor search\,\cite{malkov2018efficient}, which allows both encoders to be arbitrarily complex, as their inference is not required at runtime and user and item embeddings can be precomputed in advance.
    


Alternatively,\,\citet{rajput2023tiger} proposed generative retrieval, which uses transformer memory (parameters) as an end-to-end index. Generative retrieval often relies on representing items as sequences of semantic tokens (semantic IDs), which are typically derived by quantizing content embeddings\,\cite{rajput2023tiger, zhou2025onerec}. Quantization, however, requires high-quality content embeddings, and the sequential structure of semantic IDs increases the compute needed to process an item sequence, since each item is represented by multiple tokens.


Following\,\citet{badrinath2025pinrec}, we represent items with dense embeddings instead of relying on semantic IDs and train a two-tower model that autoregressively constructs a sequence of recommendations that is intended to maximize long-term user satisfaction. In this paper, we describe the training and evaluation of this model. Our contributions are summarized as follows:

\begin{enumerate}
    \item We propose autoregressive off-policy training of a REINFORCE\,\cite{williams1992reinforce} two-tower retriever on the public large-scale dataset Yambda-5B\,\cite{ploshkin2025yambda}. In contrast to prior work\,\cite{chen2018topk}, our autoregressive setup allows for multi-step approximations of importance weights, which empirically increase the expected reward compared to a one-step approximation.
    \item We extend the application of doubly robust estimators for sequential decision-making to the recommendation problem by training a user feedback model. To the best of our knowledge, prior work has only explored the case of contextual bandit estimators\,\cite{wang2025off} in recommendation systems.
    \item We adapt multi-step test-time scaling to an RL retriever by simulating environment dynamics with a user feedback model, whereas previous work has only considered single-step generation\,\cite{guo2026promise}.
    \item We demonstrate the superiority of an RL agent over next-item prediction and next-positive prediction baselines (which often serve as retriever training objectives) given the same amount of data, training time, and compute.
\end{enumerate}
\section{Background}
This section introduces the notation and technical background used throughout the paper.

\subsection{Sequential Decision-Making}
Sequential decision-making is often formulated as the Markov Decision Process (MDP), which is conventionally defined by a tuple \(\langle \mathcal{X}, \mathcal{A}, P_r(x, a), P(\cdot \mid x, a), P_0(x), \gamma \rangle, x \in \mathcal{X}, a \in \mathcal{A}\), where \(\mathcal{X}\) is the state space, \(\mathcal{A}\) is the action space, \(P_r(x, a)\) represents the distribution of the stochastic reward \(r(x, a)\) of taking action \(a\) in state \(x\), \(P(\cdot \mid x, a)\) is the transition probability distribution, \(P_0(x)\) is the initial state distribution, and \(\gamma \in [0, 1)\) is the discounting factor. 

\(\pi(\cdot \mid x)\) denotes a policy, which is a probability distribution over the action space conditioned on state \(x\). Since we discuss off-policy RL, we separate the behavior policy \(\pi_b(\cdot \mid x)\) (which is used for collecting training data) from the target policy \(\pi_\theta(\cdot \mid x)\) (which is the output of the training stage). In practice, we often deal with an estimation of the behavior policy \(\widehat{\pi}_b(\cdot \mid x)\), since the real distribution is rarely known.

By \(\xi = (x_0, a_0, r_{0},\ldots, x_{T_\xi-1}, a_{T_\xi-1}, r_{T_\xi-1}, x_{T_\xi})\) we denote a trajectory of length \(T_\xi\) generated by policy \(\pi\), where \(x_0 \sim P_0(x)\), \(a_t \sim \pi(\cdot \mid x_t)\), \(x_{t+1} \sim P(\cdot \mid x_t, a_t)\), and \(r_t \sim P_r(\cdot \mid x_t, a_t)\). For brevity, we denote such complex distribution as \(\xi \sim P^\pi_\xi\) and the discounted return as \(R_{t_1:t_2}(\xi)=\sum_{t=t_1}^{t_2}\gamma^{t-t_1}r_t\). Then \(R_{0:T_\xi-1}(\xi)=\sum_{t=0}^{T_\xi-1}\gamma^tr_t\) is the discounted return of the trajectory \(\xi\).

\(V^\pi(x)\) and \(Q^\pi(x, a)\) are the value functions under policy \(\pi\) for state \(x\) and state-action pair \((x, a)\), respectively. They are defined as the expected discounted return of a trajectory starting from state \(x\) (or state-action pair \((x, a)\)):
\[
V^\pi(x) = \mathbb{E}_{\xi \sim P^\pi_\xi} \left[ R_{0:T_\xi-1}(\xi) \mid x_0 = x \right],
\]
\[
Q^\pi(x, a) = \mathbb{E}_{\xi \sim P^\pi_\xi} \left[ R_{0:T_\xi-1}(\xi) \mid x_0 = x, a_0 = a \right].
\]
We evaluate a policy \(\pi\) by the expectation of discounted returns of the trajectories it generates: \(\rho^\pi = \mathbb{E}_{x_0 \sim P_0} \left[ V^\pi(x_0) \right]\).

\subsection{Off-policy REINFORCE Recommender}
The general task of an RL agent is to maximize the expected discounted return of a trajectory generated by a policy \(\pi\):
\[\pi_\theta = \arg \max_{\pi} \mathcal{J}(\pi) = \arg \max_{\pi} \mathbb{E}_{\xi \sim P^\pi_\xi} \left[ R_{0:T_\xi-1}(\xi) \right].\]

Policy gradient methods\,\cite{levine2013guided} offer a way to optimize this objective directly, for example, using the REINFORCE algorithm\,\cite{williams1992reinforce}:
\[\nabla_{\theta} \mathcal{J}(\pi_\theta) = \mathbb{E}_{\xi \sim P^{\pi_\theta}_\xi} \left[ \sum_{t=0}^{T_\xi-1} \gamma^t R_{t:T_\xi-1}(\xi) \nabla_{\theta} \log \pi_\theta(a_t \mid x_t) \right].\]

This formula assumes that the trajectories are sampled from the same policy that is being optimized. Our agent, however, does not have access to the environment and can only be trained using pre-collected trajectories generated by a behavior policy. Off-policy correction allows trajectories to be sampled from a different policy by reweighting each term by \(\omega_{t_1:t_2}(\xi)=\prod_{\tau=t_1}^{t_2}\frac{\pi_\theta(a_\tau \mid x_\tau)}{\pi_b(a_\tau \mid x_\tau)} \approx \prod_{\tau=t_1}^{t_2}\frac{\pi_\theta(a_\tau \mid x_\tau)}{\widehat{\pi}_b(a_\tau \mid x_\tau)}\)\,\cite{chen2018topk}:
\[\nabla_{\theta} \mathcal{J}(\pi_\theta) = \mathbb{E}_{\xi \sim P^{\pi_b}_\xi} \left[ \sum_{t=0}^{T_\xi-1} \gamma^t \omega_{t:T_\xi-1}(\xi) R_{t:T_\xi-1}(\xi) \nabla_{\theta} \log \pi_\theta(a_t \mid x_t) \right].\]

Unlike prior work\,\cite{chen2018topk, chen2022offpolicy}, we propose autoregressive training, which allows the entire trajectory to be processed in a single forward pass rather than requiring states to be sampled from the discounted state visitation distribution. We therefore do not introduce additional bias by ignoring the state visitation frequency difference, which cannot be computed in practice. Furthermore, our autoregressive formulation allows us to estimate \(\omega_{t:T_\xi-1}(\xi)\) precisely instead of keeping a single factor\,\cite{chen2018topk}, which avoids another bias in gradient estimation. 

Following \citet{chen2018topk}, we drop the leading factors \(\omega_{0:t-1}(\xi)\) to reduce gradient variance and weight each step only by the current and subsequent ratios. However, instead of keeping all \(T_\xi - t\) factors, we experiment with retaining up to \(K\) of them to control gradient variance.

\subsection{Off-policy Evaluation}
\label{sec:ope}
Historically, OPE estimators are divided into two branches: methods for contextual bandits and methods for sequential decision-making problems (RL). Contextual bandits may be considered as MDPs with a single step horizon: the context is sampled from the environment and assumed to be independent of the agent's actions. OPE methods for contextual bandits therefore estimate the reward per context-action pair and average them over the dataset, and only reflect the agent's ability to maximize the immediate reward. RL estimators, on the other hand, are designed for multi-step MDPs, where the environment responds to the agent's actions by sampling states from a transition distribution conditioned on the agent's action. They therefore estimate the expected return of a trajectory. Contextual bandit estimators thus can be seen as a special case of RL estimators where the trajectory has length 1 (following the introduced notations, \(\forall \xi: T_\xi = 1\)).

There are roughly two classes of approaches to OPE, which can be adapted to both contextual bandits and RL. 
\begin{enumerate}
    \item The reason we cannot replay the target policy on the dataset is that the reward is only known for the actions taken by the behavior policy. A straightforward solution is to train a model of an environment and then use it to estimate the performance of the target policy \(\pi_\theta\). This is the core idea behind the \textbf{direct method (DM)}. In the case of contextual bandits, this model is the mean reward of each context-action pair, and in RL it is either the mean reward \(\widehat{r}(x, a)\) and state transition \(\widehat{P}(\cdot \mid x, a)\), or the state (state-action) value \(\widehat{V}^{\pi_\theta}(x)\) (\(\widehat{Q}^{\pi_\theta}(x, a)\)) function directly:
    \[
    \widehat{\rho}^{\pi_\theta}_{DM} = \frac{1}{N} \sum_{i=1}^N R_{0:T_{\xi_i}-1}(\xi_i) = \frac{1}{N} \sum_{i=1}^N \widehat{V}^{\pi_\theta}(x_i),
    \]
    where \(\xi_i \sim \widehat{P}^{\pi_\theta}_\xi\) (\(\widehat{P}^{\pi_\theta}_\xi\) is induced by \(P_0(x)\), \(\widehat{P}(\cdot \mid x, a)\), and \(\widehat{r}(x, a)\)) and \(x_i \sim P_0(x)\).

    Given that the environment model is trained on data produced by another (behavior) policy, it may explore areas irrelevant to the target policy, which can lead to uncontrolled bias but generally low variance\,\cite{dudik2011doubly, farajtabar2018more}.
    \item Another method, called \textbf{inverse propensity scoring (IPS)}, is based on the idea of estimating the expected reward by reweighting rewards observed under the behavior policy via importance sampling (IS). Throughout this paper, we will use its step-wise variant:
    \[
    \widehat{\rho}_{\text{step-IS}}^{\pi_\theta} = \frac{1}{N} \sum_{i=1}^N \sum_{t=0}^{T_{\xi_i}-1}\gamma^t \omega^{(i)}_{0:t}r^{(i)}_t,
    \]
    where \(\omega^{(i)}_{t_1:t_2} = \prod_{\tau=t_1}^{t_2}\frac{\pi_\theta(a_\tau \mid x_\tau)}{\pi_b(a_\tau \mid x_\tau)}\) is the cumulative importance weight for the \(t_1\)-th to \(t_2\)-th step of the \(i\)-th trajectory \(\xi_i \sim P^{\pi_b}_\xi\), and \(r^{(i)}_t\) is the observed reward.

    When the behavior policy is known, the IPS estimator is unbiased under the absolute continuity assumption (any state-action pair that has zero probability under the behavior policy has zero probability under the target policy as well). However, it often suffers from severe variance due to the unbounded importance weights. In practice, the behavior policy is often unknown and must be estimated, which introduces bias into the IPS estimator.

    Note that \citet{jiang2016doubly} and \citet{farajtabar2018more} suggest using another variant of the IPS estimator (a step-wise weighted importance sampling estimator) which is biased but reduces variance even further. Our case differs in that the collected behavioral trajectories have different lengths, leading to uneven weighting of trajectory steps and introducing noise to the estimate. We therefore use a step-wise importance sampling estimator (step-IS).
\end{enumerate}

To have the best of both worlds, DM and IPS were integrated within the \textbf{doubly robust (DR)} estimator, first for statistics\,\cite{cassel1976some}, then for contextual bandits\,\cite{dudik2011doubly}, and finally for RL\,\cite{jiang2016doubly, farajtabar2018more}:
\begin{align*}
    \widehat{\rho}_{\text{DR}}^{\pi_\theta} &= \frac{1}{N} \sum_{i=1}^N \sum_{t=0}^{T_{\xi_i}-1}\Big[\gamma^t\omega_{0:t}^{(i)}r^{(i)}_t \\
    &- \gamma^t\big(\omega_{0:t}^{(i)}\widehat{Q}^{\pi_\theta}(x^{(i)}_t,a^{(i)}_t)-\omega_{0:t-1}^{(i)}\widehat{V}^{\pi_\theta}(x^{(i)}_t)\big)\Big]\footnotemark.
\end{align*}
\footnotetext{Here, \(\widehat{Q}^{\pi_\theta}(x_t,a_t)\) and \(\widehat{V}^{\pi_\theta}(x_t)\) denote the expected return of the trajectory suffix, which we treat as a valid trajectory of lenth \(T_\xi - t\).}
\citet{dudik2011doubly} have shown that the DR estimator for contextual bandits is roughly unbiased when either DM or IPS has low bias (i.e., when either the behavior policy estimate or the reward model is accurate). \citet{farajtabar2018more} have later extended the bias analysis to the case of RL, proving the bias to be:
\begin{equation}
Bias(\widehat{\rho}_{\text{DR}}^{\pi_\theta}) = | \mathbb{E}_{P^{\pi_\theta}_\xi} \left[ \sum_{t=0}^{T_\xi-1}\gamma^t\lambda_{0:t-1}(\xi)\delta_{0:t}(\xi)\Delta(x_t,a_t) \right]|,
\label{eq:bias}
\end{equation}
where \(\lambda_{0:t}(\xi)=\prod_{\tau=0}^t\frac{\pi_b(a_\tau \mid x_\tau)}{\widehat{\pi}_b(a_\tau \mid x_\tau)}\), \(\delta_{0:t}(\xi)=1-\lambda_{0:t}(\xi)\), and \(\Delta(x,a)=\widehat{Q}^{\pi_\theta}(x, a)-Q^{\pi_\theta}(x,a)\). Note that similar to contextual bandits\,\cite{dudik2011doubly}, when the behavior policy is known (\(\widehat{\pi}_b=\pi_b\), implying that \(\delta_{0:t}(\xi)=0\)) or the DM estimator is unbiased (\(\widehat{Q}^{\pi_\theta}(x, a) = Q^{\pi_\theta}(x,a)\)), the bias of the DR estimator is zero.

To the best of our knowledge, in recommender systems, prior work appears to have explored DR only for contextual bandits\,\cite{wang2025off}.
\section{Method}

\paragraph{Recommendation as RL problem}
We formulate music recommendation as a sequential decision-making problem. The setting corresponds to a streaming feed, where at each step, the user interacts with a recommended track, provides feedback, and receives the next recommendation.

In our notation, at each step \(t\), the state \(x_t\) is the user history up to this step, and the action \(a_{t+1} \in \mathcal{A}\) is the track (item) selected from the catalog by a recommender. Feedback includes a binary like indicator \(\mathbbm{1} \left\{ \texttt{like}_t \right\}\) and the listened \(\texttt{ratio}_t \in [0, 1]\). We define the immediate reward as \(r_t = r(x_t,a_t) = \mathbbm{1} \left\{ \texttt{like}_t \right\} + \frac{1}{10} \texttt{ratio}_t\), \(r_t \in [0, 1.1]\). The next state \(x_{t+1}\) is obtained by appending the recommended track and the observed feedback to the user history.

We treat sequences of consecutive non-organic interactions (interactions with recommended tracks) as trajectories (sessions), since only recommendation events are controlled by the policy. Organic events (i.e., user searching for a specific track) are included as part of the user state, but are not treated as policy actions. Our task is to maximize the expected discounted per-trajectory return, where we set \(\gamma=0.9\).

\subsection{Parameterization}
\paragraph{Behavior policy}
We first train a two-tower model \(f_b\) that estimates the behavior probabilities \(\hat{\pi}_b\) needed to calculate importance weights. The user tower is a causal transformer decoder that inputs a chronologically ordered sequence of events, where each event consists of the context (whether it is an organic or recommendation event), item information (track ID and duration), and user feedback (binary like and quantized listened ratio). The item tower is an MLP that encodes item information. The user-candidate similarity is modeled by the dot-product between the user embedding (behavior transformer hidden state corresponding to input state \(x\)) \(e_{b,x}\) and the candidate embedding \(e_{b,c}\): \(f_b(x, c) = \langle e_{b,x}, e_{b,c} \rangle\).

We use the next-item prediction objective with the logQ correction\,\cite{yi2019sampling}:
\[
\mathcal{L}_{\mathrm{NIP}}(x, c) = -\log \frac{\exp \left\{f_b(x, c)\right\}}{\exp \left\{f_b(x, c)\right\} + \sum_{n \in \mathcal{N}} \exp \left\{f_b(x, n)-\log Q(n)\right\}},
\]
where \(\mathcal{N}\) is the sampled negative set and \(Q(n)\) is the negative-sampling probability. Since only non-organic interactions are controlled by the policy, loss is calculated only for non-organic targets in the user sequence (if any). Note that causal masking enables autoregressive training, in which all user states in the batch are encoded with a single transformer forward pass.

Behavior probability is then calculated as follows:
\begin{equation}
    \label{eq:probability}
    \hat{\pi}_b(a \mid x_t) = \frac{\exp \left\{f_b(x_t, a)\right\}}{\sum_{a^{\prime} \in \mathcal{A}}\exp \left\{f_b(x_t, a^{\prime})\right\}}.
\end{equation}

\paragraph{Target policy}
The target policy is parameterized with exactly the same architecture as the behavior policy. Furthermore, we initialize the target policy model \(f_\theta\) from the trained behavior policy model \(f_b\). 

\begin{figure}[t]
    \centering
    \includegraphics[width=1\linewidth]{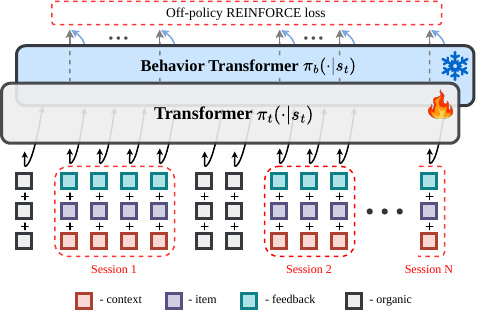}
    \caption{Autoregressive off-policy REINFORCE training. To calculate importance weights in REINFORCE loss, we estimate the behavior policy probabilities with a frozen transformer. The target policy is also parametrized with a transformer that is trained using REINFORCE loss. Consecutive non-organic events form recommendation sessions (trajectories), serving as targets in REINFORCE. Organic events are only used as a part of user history (state).}
    \label{fig:offpolicy_reinforce_pipeline}
\end{figure}

The target policy model is trained autoregressively with an off-policy REINFORCE objective over non-organic sessions, while the behavior policy model is frozen. Mini-batch processing requires a single forward pass for both target and behavior policy transformers. Note that each user sequence may contain multiple non-organic sessions (trajectories). \autoref{fig:offpolicy_reinforce_pipeline} visualizes a single training step. Unlike prior work, which constructed an independent training example for each state-action pair\,\cite{chen2018topk}, we process the whole user sequence in a single forward pass.

For a trajectory \(\xi\), we approximate each step's importance weight with at most \(K\) policy-ratio factors:
\[
\omega_{t:t+K-1}(\xi) = \prod_{\tau=t}^{\min(t+K-1,\,T_\xi-1)} \frac{\pi_\theta(a_\tau \mid x_\tau)}{\widehat{\pi}_b(a_\tau \mid x_\tau)}.
\]
Weight capping \(\bar{\omega}_{t:t+K-1}(\xi) = \min\left(c,\,\omega_{t:t+K-1}(\xi)\right)\), where \(c\) is the clipping threshold, is applied to reduce the variance of gradients. \(\pi_\theta\) is calculated similarly to \autoref{eq:probability}, but using target policy model \(f_\theta\) instead of \(f_b\).

Let \(\mathcal{B}\) be a mini-batch of user sequences, and let \(\mathcal{S}(i)\) denote the set of non-organic sessions (trajectories) contained in the \(i\)-th user sequence. The off-policy REINFORCE gradient is then estimated as:
\begin{align*}
    &\widehat{\nabla_{\theta}\mathcal{J}}(\mathcal{B}) = \frac{1}{\sum_{i = 1}^{|\mathcal{B}|} \sum_{\xi\in\mathcal{S}(i)} \sum_{t=0}^{T_\xi-1} \gamma^t \bar{\omega}_{t:t+K-1}(\xi)} \times {} \\
    &\quad \biggl( \sum_{u\in\mathcal{B}} \sum_{\xi\in\mathcal{S}(u)} \sum_{t=0}^{T_\xi-1} \gamma^t \bar{\omega}_{t:t+K-1}(\xi) R_{t:T_\xi-1}(\xi)\nabla_\theta \log \pi_\theta(a_t \mid x_t) \biggr).
\end{align*}
Note that a single-step estimation of the importance weight proposed by \citet{chen2018topk} is a special case of the above when \(K=1\) and only the current action contributes to the weight. 
In contrast, our multi-step approximation accounts for several future policy-ratio factors within the same trajectory while avoiding the variance of full-trajectory importance sampling.

\paragraph{User feedback model}
To perform model-based evaluation, as well as test-time scaling, we need to imitate user decisions. Given the current user state \(x_t\) and a candidate action \(a_t\), we train the model that predicts the feedback the user provides after interacting with the recommended track:
\[
\hat f_t \sim \widehat{P}_{\text{feedback}}(\cdot \mid x_t,a_t) = \widehat{P}_{\text{like}}(\cdot \mid x_t,a_t) \times \widehat{P}_{\text{ratio}}(\cdot \mid x_t,a_t).
\]
Note that instead of predicting the reward \(r_t\) directly, we independently predict the listened ratio and the probability of the user liking this track. This is because our policy model \(f_\theta\) inputs the feedback from each past interaction (which consists of the binary like and the listened ratio), and in order to call this model iteratively to generate the entire trajectory, we have to predict feedback for each trajectory element. 

\begin{figure}[t]
    \centering
    \includegraphics[width=1\linewidth]{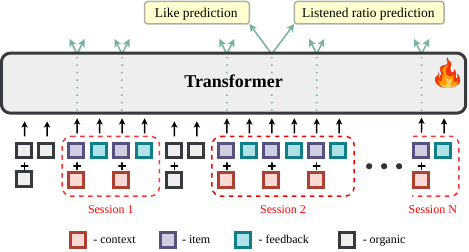}
    \caption{User feedback model. Given the user history and candidate recommendation events, transformer predicts the feedback that the user would provide for each target item. The model has two heads for like prediction and listened ratio prediction, which are then used to compute the predicted reward for model-based evaluation.}
    \label{fig:user_preference_model_pipeline}
\end{figure}

The user feedback model is parameterized by a causal transformer. We want this model to autoregressively predict feedback for each action (non-organic event) in the sequence while taking advantage of "seeing" feedback for all past actions. For these two reasons, we organize the input sequence as action-feedback pairs (each input event yields two input tokens). See  \autoref{fig:user_preference_model_pipeline} for the visualization.

We take transformer hidden states that correspond to action inputs and project them into scalar like \(\ell_{\text{like}}(x_t,a_t)\) and ratio \(\ell_{\text{ratio}}(x_t,a_t)\) logits, that are used to calculate the corresponding losses:
\[
\mathcal{L}_\text{like}(\ell_{\text{like}}(x_t,a_t), \texttt{like}_t) = \text{BCEWithLogits}(\ell_{\text{like}}(x_t,a_t), \mathbbm{1} \left\{ \texttt{like}_t \right\}),
\]
\[
\mathcal{L}_\text{ratio}(\ell_{\text{ratio}}(x_t,a_t), \texttt{ratio}_t) = \text{BCEWithLogits}(\ell_{\text{ratio}}(x_t,a_t), \texttt{ratio}_t).
\]
The final loss is a weighted sum of the above:
\begin{align*}
    \mathcal{L} = \sum_t \biggl( &\frac{2}{10} \mathcal{L}_\text{like}(\ell_{\text{like}}(x_t,a_t), \texttt{like}_t) \\ &+ \frac{8}{10} \mathcal{L}_\text{ratio}(\ell_{\text{ratio}}(x_t,a_t), \texttt{ratio}_t)\biggr).
\end{align*}

Predicted feedback is then calculated as \(\widehat{\texttt{like}}_t = \sigma(\ell_{\text{like}}(x_t,a_t))\) and \(\widehat{\texttt{ratio}}_t = \sigma(\ell_{\text{ratio}}(x_t,a_t))\). The expected reward is estimated as \(\widehat{r}_t = \mathbbm{1} \left\{ \widehat{\texttt{like}}_t \right\} + \frac{1}{10} \widehat{\texttt{ratio}}_t\).

In our experiments, we also initialize the user feedback model from the trained behavior policy model \(f_b\).

\subsection{Multi-Step Test-Time Scaling}

\begin{figure}[t]
    \centering
    \includegraphics[width=1\linewidth]{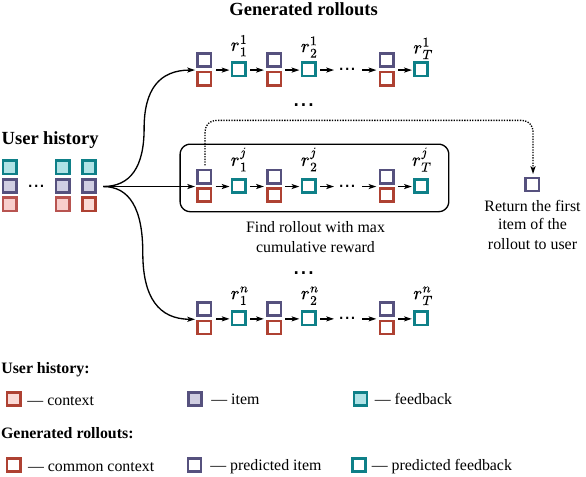}
    \caption{Multi-step test-time scaling. Given the user history, we sample \(N\) roll-outs of length \(T\). Each roll-out is scored by its predicted discounted return under the learned feedback model. The procedure returns only the first item from the roll-out with the highest cumulative reward.}
    \label{fig:test_time_scaling_pipeline}
\end{figure}

At inference time, it is possible to spend additional compute to evaluate several possible future trajectories before selecting a recommendation.

Firstly, we are given the current user state \(x_0\). We build the \(T\)-step trajectory using the following iterative procedure, which at each step \(t = 0, \ldots, T-1\) returns the action \(a_t\). For the \(t\)-th step:
\begin{enumerate}
    \item \(N\) roll-outs are initialized with the current state: \(x_t^{(j)} = x_t, j = 0, \ldots, N - 1\);
    \item each roll-out is continued up to length \(T\). Let \(t^\prime =t\), then for the \(j\)-th roll-out:
    \begin{enumerate}
        \item the trained target policy samples an action \(a^{(j)}_{t^\prime} \sim \pi_\theta(\cdot \mid x^{(j)}_{t^\prime})\);
        \item the trained user feedback model predicts the feedback (\(\widehat{\texttt{like}}_{t^\prime}^{(j)}\) and \(\widehat{\texttt{ratio}}_{t^\prime}^{(j)}\)) for the sampled action \(a^{(j)}_{t^\prime}\), defining its reward \(\widehat{r}_{t^\prime}^{(j)}\);
        \item \(t^{\prime}\) is incremented, and the generation process is repeated while \(t^\prime \leq T-1\);
    \end{enumerate}
    \item each roll-out is then scored by its predicted discounted return, and the best roll-out, \(j^{\star}\), is selected;
    \item the new state \(x_{t+1}\) is obtained by appending the predicted action \(a^{(j^\star)}_t\) and feedback \(\widehat{\texttt{like}}_t^{(j^\star)}\), \(\widehat{\texttt{ratio}}_t^{(j^\star)}\) at step \(t\) to the previous state \(x_t\);
    \item \(t\) is incremented, and the procedure is repeated while \(t \leq T-1\).
\end{enumerate}
A single iteration of this procedure is visualized in \autoref{fig:test_time_scaling_pipeline}.

This method is inspired by the general idea of estimating the value of intermediate decisions via Monte-Carlo continuations, as in recent step-level value estimation and test-time scaling approaches\,\cite{wang2024mathshepherd, snell2024scaling}. Unlike methods that train a separate process reward model, we explicitly perform model-based roll-outs at inference time and use their predicted returns to choose the next recommendation. To make this procedure efficient, we reuse the KV-cache of both the target policy and the user feedback model across roll-out steps.

\subsection{Supervised Baselines}
We compare the REINFORCE recommender with two supervised baselines that are often used for candidate generation in practice. Similar to the REINFORCE recommender, they both have the same architecture and are initialized from the trained behavior policy model \(f_b\).
\begin{enumerate}
    \item \textbf{NIP-CE} continues next-item prediction for non-organic targets.
    \item \textbf{Positive-CE} follows a common candidate-generation setup in which the model is trained only on positive interactions. The interaction is considered positive when \(r(x_t,a_t)\geq 0.1\) (i.e., it is either liked or listened to in full by a user).
\end{enumerate}

Both baselines are trained on the same amount of data as the REINFORCE recommender, requiring the same training time and compute.

\subsection{Implementation Details}
To encode the high-cardinality item ID feature, we use multi-hash technique\,\cite{svenstrup2017hash}, which applies multiple hash functions to the same item ID, retrieves several embedding table entries, and combines them by summation. The embedding table contains 524,288 rows. The listened ratio is quantized into 12 intervals, and each interval is assigned a unique embedding. Track duration is quantized into 7 intervals, which are then mapped to unique embeddings, similar to the listened ratio.

For our transformer models, we set the following hyperparameters: \(\texttt{num\_layers}=4\), \(\texttt{hidden\_dim}=256\), \(\texttt{num\_heads}=4\). All models are trained with AdamW optimizer.

The behavior policy is trained with \(\texttt{learning\_rate}=10^{-3}\) for 10 epochs, using sampled softmax with 16,384 in-batch negative examples. The REINFORCE recommender, as well as the supervised baselines, is trained for 10 epochs with \(\texttt{learning\_rate}=10^{-4}\). Their training budgets are matched in terms of data, training time, and compute to ensure a fair comparison. For off-policy RL training, we use up to 10 importance-weight factors and clip the weights at 0.5. The user feedback model is trained for 5 epochs with \(\texttt{learning\_rate}=10^{-3}\).

In all test-time scaling experiments, we use a roll-out length of \(T=8\). We generate candidate roll-outs using top-\(k\) sampling with \(k=4\).

\section{Experiments}
We formulate the following research questions:
\begin{itemize}
    \item \textbf{RQ1:} Can we improve long-term user satisfaction with policy gradient optimization without sacrificing performance based on standard retrieval metrics and without requiring additional data, training time, or compute?
    \item \textbf{RQ2:} Following the LLM experience, can we increase retrieval quality with multi-step test-time scaling by taking advantage of user feedback model?
\end{itemize}

\subsection{Experimental Setup}
\paragraph{Data}
We conduct experiments on Yambda-5B\,\cite{ploshkin2025yambda}, a large-scale public dataset. We use its largest version, with five billion user interactions, since this scale makes it similar to real-world production scenarios. We believe that dataset scale is particularly important because conclusions made on tiny datasets may not necessarily transfer to an industrial setting\,\cite{dacrema2021troubling}.

To avoid data leakage\,\cite{ji2023critical, meng2020exploring}, we use a temporal split: the last week is held out for testing, while the remaining data is used for training. Each user history is split into chunks of 512 consecutive events.

\paragraph{Evaluation metrics}
We divide metrics into two classes: recall-based retrieval metrics and off-policy evaluation metrics.

\textbf{Recall-based metrics} are used to ensure that RL optimization does not derail retrieval quality. These metrics, however, are not valid estimates of the long-term effect, since they only evaluate the policy at a finite set of points generated by a different policy. 

We calculate recall using an evaluation index \(\mathcal{I}\) that contains the top 500,000 most popular tracks from the training period. If the target track is not present in \(\mathcal{I}\), we evaluate recall as if the target track had been added to the index, i.e., using the expanded index \(\mathcal{I} \cup \{a\}\) of size 500,001. This ensures that out-of-index targets are still treated as valid candidates during evaluation.

For a user state \(x\), a target action \(a\), and an evaluation index \(\mathcal{I}\), let \(\text{rank}^{\pi}_{\mathcal{I}}(a \mid x)\) denote the rank of this action under policy \(\pi\). Then recall@K over a target set \(\mathcal{D}\) is defined as
\[
\text{r}^{\pi}_{\mathcal{D}}\text{@K} = \frac{1}{|\mathcal{D}|} \sum_{(x,a)\in \mathcal{D}} \mathbbm{1} \left\{ \text{rank}^{\pi}_{\mathcal{I} \cup \{a\}}(a \mid x) \leq K \right\}.
\]

We report recall@100 on three target sets: 
\begin{enumerate}
    \item all non-organic state-action pairs \(\mathcal{D}_{\text{all}}\). Recall on \(\mathcal{D}_{\text{all}}\) targets measures the ability to clone behavior policy;
    \item positive pairs \(\mathcal{D}_{\text{pos}}=\{(x_t,a_t)\in\mathcal{D}_{\mathrm{all}}: r_t\geq 0.1\}\). Recall on \(\mathcal{D}_{\text{pos}}\) targets corresponds to the standard evaluation of a candidate generator on positive interactions;
    \item return-based slices \(\mathcal{D}_{pA\text{--}B} = \{(x_t,a_t)\in\mathcal{D}_{\mathrm{all}}: R_{t:T_\xi-1}(\xi) \in [q_A,q_B]\}\), where \(q_A\) and \(q_B\) are the \(A\)-th and \(B\)-th percentiles of the return distribution. In our experiments, we use the low-return slice \(p00\text{--}05\) and the high-return slice \(p95\text{--}100\). \(\text{r}^{\pi}_{\mathcal{D}_{pA\text{--}B}}\text{@K}\) is used as a proxy for analyzing which actions the model tends to reproduce: those associated with low or high future session-level return. The desired behavior is to avoid degrading recall on high-return actions while reducing the likelihood of reproducing low-return actions.
\end{enumerate}

To estimate long-term reward, we use \textbf{OPE estimators} introduced in \autoref{sec:ope}: Step-IS, DM, and DR. Step-IS applies importance sampling correction with clipping at 10. DM evaluates the policy using the learned user feedback model. \(V^\pi(x)\) and \(Q^\pi(x, a)\) functions are estimated with Monte-Carlo roll-outs using greedy decoding.

\subsection{Long-Term User Satisfaction (RQ1)}
\begin{table}[t]
  \centering
  \caption{Recall-based comparison.}
  \label{tab:recall_all_targets}
  \Description{Comparison of NIP-CE, Positive-CE, and Off-Policy RL models using recall@100 across all, positive, low-reward, and high-reward evaluation targets.}
  \small
  \setlength{\tabcolsep}{6pt}
  \begin{tabular}{lcccc}
    \toprule
    Model
    & \(\mathcal{D}_{\text{all}}\)
    & \(\mathcal{D}_{\text{pos}}\)
    & \(\mathcal{D}_{p00\text{--}05}\)
    & \(\mathcal{D}_{p95\text{--}100}\) \\
    \midrule
    NIP-CE
      & \textbf{0.3153}
      & \underline{0.3217}
      & \textbf{0.2814}
      & \textbf{0.2679} \\
    Positive-CE
      & 0.3029
      & \textbf{0.3298}
      & 0.2634
      & \underline{0.2662} \\
    REINFORCE
      & \underline{0.3086}
      & 0.3174
      & \underline{0.2653}
      & 0.2655 \\
    \bottomrule
  \end{tabular}
\end{table}

\paragraph{Evaluation based on standard retrieval metrics}
\autoref{tab:recall_all_targets} compares the REINFORCE recommender with supervised baselines based on recall@100 across the four different target sets discussed above. Several observations follow from this experiment.

First, the best performance across \(\mathcal{D}_{\text{all}}\) and \(\mathcal{D}_{\text{pos}}\) is achieved by NIP-CE and Positive-CE baselines, respectively. This is expected, given that these models have been directly optimized for these target sets. Fortunately, the performance gap between the baselines and the REINFORCE recommender is quite modest.

Second, there is a clear difference between NIP-CE and the other two models in \(\text{r}^{\pi}_{\mathcal{D}_{p00\text{--}05}}\text{@100}\), suggesting that the Positive-CE and the REINFORCE recommender are less prone to retrieving actions considered negative in long-term. This observation is a natural consequence of the RL objective: the model should be less likely to reproduce actions associated with low future reward. Based on \(\mathcal{D}_{p95\text{--}100}\), all models seem to perform on-par.

\paragraph{Off-policy evaluation.}
\begin{table}[t]
  \centering
  \caption{User feedback model evaluation.}
  \label{tab:feedback_world_model}
  \Description{Evaluation of the user feedback model used to predict like probability and played ratio. The reward is computed deterministically from the predicted feedback signals.}
  \small
  \setlength{\tabcolsep}{4pt}
  \begin{tabular}{lccc}
    \toprule
    Model & Ratio MAE & Like PR-AUC & Reward MAE \\
    \midrule
    User Feedback Model & 0.1771 & 0.2432 & 0.0340 \\
    \bottomrule
  \end{tabular}
\end{table}

We first evaluate the user feedback model, which is a key component of off-policy evaluation. \autoref{tab:feedback_world_model} shows the prediction quality of the user feedback model. The listened ratio, which is an important signal since it serves as a dense component of the reward, is predicted reasonably accurately. Like prediction is more challenging due to the high sparsity of likes. Note that the fraction of liked tracks in the data is 1.13\%, resulting in a PR-AUC of random guessing of around 0.0113. Nevertheless, the reward MAE remains low, which enables the model to be used for model-based and doubly robust evaluation.

\begin{table}[t]
  \centering
  \caption{Off-policy evaluation comparison. Behavior return is calculated directly from the test data to measure the scale of the expected return.}
  \label{tab:ope_metrics}
  \Description{Comparison of logged policy reward, NIP-CE, Positive-CE, and Off-Policy RL models using Step-IS, direct method, and double-robust off-policy evaluation metrics.}
  \small
  \setlength{\tabcolsep}{4pt}
  \begin{tabular}{lcccc}
    \toprule
    Model
    & Behavior Return
    & Step-IS
    & DM
    & DR \\
    \midrule
    Behavior Policy  & 0.5855 & -- & -- & -- \\
    NIP-CE           & -- & 1.3065 & 0.5279 & 0.6046 \\
    Positive-CE      & -- & \underline{1.3317} & \underline{0.5384} & \underline{0.6184} \\
    REINFORCE    & -- & \textbf{1.4677} & \textbf{0.5395} & \textbf{0.6306} \\
    \bottomrule
  \end{tabular}
\end{table}

The OPE results are presented in \autoref{tab:ope_metrics}. 

First, they confirm that the REINFORCE recommender achieves the best results across all OPE metrics. This is the main experimental evidence that the explicit optimization of the discounted per-session return improves long-term utility compared to the supervised objectives.

Second, Positive-CE also outperforms NIP-CE across all metrics. This indicates that optimizing immediate positive feedback can partially improve session-level metrics. However, the gains of Positive-CE are smaller than those of the REINFORCE recommender, illustrating the difference between the local optimization of positive actions and the explicit optimization of cumulative per-session return.

An additional insight comes from comparing the Step-IS, DM, and DR estimate values with the mean return of the true behavior policy. The Step-IS estimate differs substantially from the behavior return scale, whereas DM matches it better. This suggests that, in our setting, DM has lower empirical bias than Step-IS. The discrepancy in Step-IS can be explained by importance-weight clipping and imperfect estimation of the true behavior policy. More interestingly, DR produces the estimate closest to the behavior return scale, which is consistent with its design: it combines historical samples and model-based estimates and can partially offset the biases of both components (See \autoref{eq:bias}).

\subsection{Test-Time Scaling (RQ2)}
\begin{figure}[t]
    \centering
    \includegraphics[width=1\linewidth]{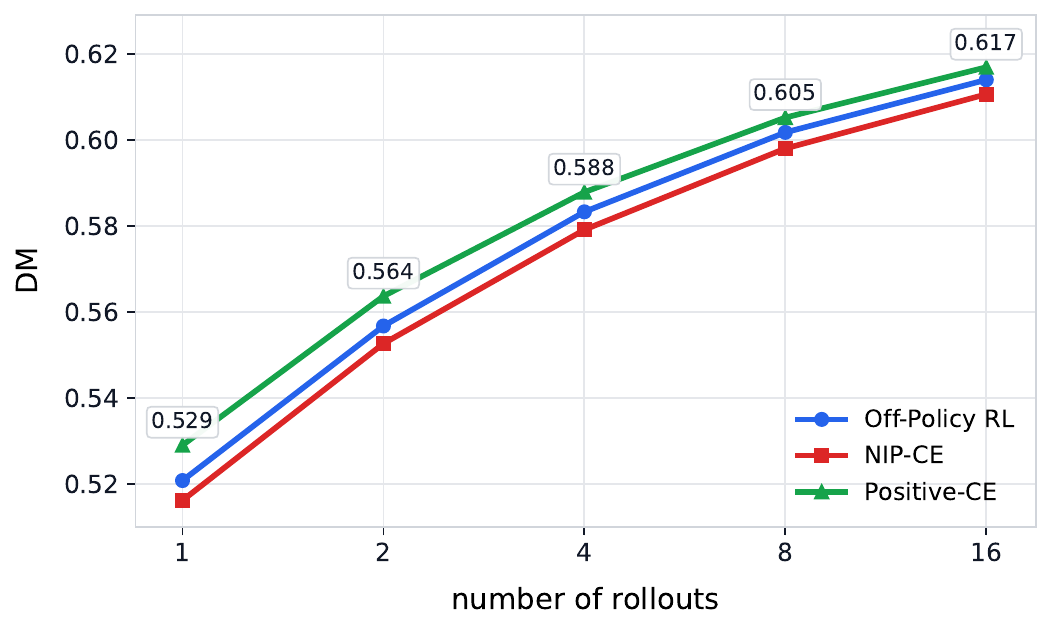}
    \caption{DM estimate versus the number of inference-time roll-outs in multi-step test-time scaling with top-\(k\) generation (\(k=4\)) and roll-out length \(T=8\). Increasing the roll-out budget improves the predicted long-term return across all models.}
    \label{fig:test_time_scaling_plot}
\end{figure}

The goal of the test-time scaling experiments is to determine whether additional roll-out selection at inference, using the learned user feedback model, can improve quality relative to the same model. In other words, we assess whether increasing inference-time compute allows the model to discover actions with higher predicted session-level reward.

At the same time, it is important to note the limitation of this evaluation. Since both roll-out selection and subsequent evaluation are performed using the same user feedback model, these results do not constitute independent evidence that the real policy quality improves. They only show that the proposed procedure can improve the model-based reward estimate. Nevertheless, this result is non-trivial: under ineffective scaling or an insufficiently informative user feedback model, the dependence of quality on the number of roll-outs could have degenerated into an almost constant curve.

Another interesting observation is that such inference-time selection allows session-level reward to be captured even for models that were trained with immediate-feedback objectives. In particular, \autoref{fig:test_time_scaling_plot} shows that the Positive-CE baseline becomes comparable to Off-Policy RL (REINFORCE) and even slightly outperforms it in terms of the DM estimate. This occurs because, at inference time, we explicitly choose the action that leads to the highest predicted cumulative reward among the considered roll-outs. The NIP-CE baseline performs the worst.

Increasing the number of parallel roll-outs yields higher DM estimates for all models. This shows that test-time scaling can provide an additional mechanism for improving model-based session-level utility without changing the parameters of the policy itself.

\subsection{Ablation Study}
\begin{table}[t]
  \centering
  \caption{Ablation on off-policy correction in REINFORCE training.}
  \label{tab:offpolicy_correction_ablation}
  \Description{Ablation study of off-policy correction weights in REINFORCE training. The variant without correction removes importance weights from the policy-gradient loss. We report off-policy evaluation metrics and recall@100 on high-reward target slices.}
  \small
  \setlength{\tabcolsep}{4pt}
  \begin{tabular}{lcccc}
    \toprule
    Model
    & Step-IS
    & DM
    & DR
    & \(\text{r}^{\pi}_{\mathcal{D}_{p95\text{--}100}}\text{@100}\) \\
    \midrule
    off-policy correction
      & \textbf{1.4677}
      & \textbf{0.5395}
      & \textbf{0.6306}
      & 0.2655 \\
    w/o off-policy correction
      & 1.2900
      & 0.5315
      & 0.6130
      & \textbf{0.2689} \\
    \bottomrule
  \end{tabular}
\end{table}

\paragraph{Off-policy correction.}
\autoref{tab:offpolicy_correction_ablation} demonstrates the importance of off-policy correction when training REINFORCE on data collected under a different policy. Without importance weights, the model ignores the distribution shift between the behavior policy and the target policy. This leads to worse Step-IS, DM, and DR values.

At the same time, the variant without correction achieves higher \(\text{r}^{\pi}_{\mathcal{D}_{p95\text{--}100}}\text{@100}\). This result highlights the limitations of recall-based proxy metrics: a stronger ability to reproduce behavior high-return actions does not necessarily imply a better estimate of the new policy in terms of off-policy value. Therefore, such recall slices are useful as diagnostic checks, but should not be treated as the primary metric of session-level effectiveness.

\begin{table}[t]
  \centering
  \caption{Ablation on the number of importance-weight factors in off-policy REINFORCE training.}
  \label{tab:num_factors_ablation}
  \Description{Ablation study of the number of importance-weight factors used for off-policy correction, evaluated with off-policy metrics and recall@100 on all targets.}
  \small
  \setlength{\tabcolsep}{5pt}
  \begin{tabular}{ccccc}
    \toprule
    \# factors & Step-IS & DM & DR & \(\text{r}^{\pi}_{\mathcal{D}_{\text{all}}}\text{@100}\)\\
    \midrule
    1
      & 1.4109
      & 0.5385
      & 0.6250
      & \textbf{0.3106} \\
    5
      & \textbf{1.4728}
      & 0.5390
      & 0.6292
      & 0.3092 \\
    10
      & \underline{1.4677}
      & \textbf{0.5395}
      & \textbf{0.6306}
      & 0.3086 \\
    15
      & 1.4536
      & \underline{0.5395}
      & 0.6295
      & 0.3085 \\
    \bottomrule
  \end{tabular}
\end{table}

\paragraph{Multi-step importance correction.}
\autoref{tab:num_factors_ablation} shows that using only one factor in the correction weight is insufficient for achieving the best performance. This differs from the setting of\,\citet{chen2018topk}, where only the first factor is used. The best result is achieved with 10 factors. Increasing the number of factors beyond 10 does not capture additional gains and may increase variance. Therefore, we use 10 factors in the main experiments.

\begin{table}[t]
  \centering
  \caption{Ablation on importance-weight capping in off-policy REINFORCE training.}
  \label{tab:importance_weight_capping_ablation}
  \Description{Ablation study of the importance-weight cap used for off-policy correction during training, evaluated with off-policy metrics and recall@100 on all targets.}
  \small
  \setlength{\tabcolsep}{5pt}
  \begin{tabular}{ccccc}
    \toprule
    IW cap & Step-IS & DM & DR & \(\text{r}^{\pi}_{\mathcal{D}_{\text{all}}}\text{@100}\) \\
    \midrule
    0.50
      & \textbf{1.4677}
      & \textbf{0.5395}
      & \textbf{0.6306}
      & \textbf{0.3086} \\
    1.00
      & \underline{1.4558}
      & \underline{0.5388}
      & 0.6295
      & \underline{0.3051} \\
    5.00
      & 1.2700
      & 0.5364
      & 0.6249
      & 0.2895 \\
    10.00
      & 1.1611
      & 0.5352
      & \underline{0.6306}
      & 0.2810 \\
    \bottomrule
  \end{tabular}
\end{table}

\paragraph{Importance-weight capping.}
\autoref{tab:importance_weight_capping_ablation} shows the effect of the clipping value for importance weights. A higher cap retains larger correction weights, but also increases gradient variance and may lead to degradation in retrieval quality. In our experiments, a cap of 0.5 achieves the best balance between improving off-policy metrics and maintaining training stability. Therefore, we use it in the main configuration.

\section{Discussion}
In our current implementation, the behavior policy, target policy, and user feedback model are trained as separate models. This design makes the experimental setup explicit. However, this is not a fundamental limitation. A practical alternative is a shared backbone with separate behavior, target, and feedback heads. Our preliminary experiments suggest that this parameterization preserves both recall-based and off-policy metrics while simplifying training and deployment. Such a unified architecture would also make test-time scaling more practical. A single model could generate candidate trajectories and estimate their expected utility, turning test-time scaling into an inference-time reranking stage without a separate reward-model stack. This provides a path toward a unified recommender rather than a classical cascaded recommendation system.

The main limitation of our study is that all results are obtained in an offline setting. Although we use OPE estimators, offline evaluation cannot fully replace online A/B testing.
\section{Conclusion and Future Work}
In this work, we studied off-policy session-level optimization for large-scale two-tower retrieval. We formulated recommendation as session-level sequential decision-making and trained an autoregressive policy with off-policy REINFORCE on the pre-collected data. Experiments on Yambda-5B show that our method
achieves higher off-policy estimates  of per-session discounted return compared to both next-item prediction and positive-only supervised training, while largely preserving retrieval quality. We also showed that the user feedback model enables off-policy evaluation estimations and multi-step test-time scaling, where additional inference-time roll-outs improve model-based session-level return estimates without updating the policy.

As the next step, we plan to study the correlation between the offline metrics and online A/B test results in a production candidate generation setting for a short-video platform. Another important direction is model scaling and analyzing how increased model capacity and inference-time compute affect session-level reward optimization in recommendation.

\bibliographystyle{ACM-Reference-Format}
\bibliography{refs}

\end{document}